\documentclass[aps,pre,twocolumn,superscriptaddress,amsmath,amsfonts,showpacs]{revtex4}
\usepackage{graphicx}

\begin{document}

\title{Response of discrete nonlinear systems with many degrees
       of freedom}

\author{Yaron Bromberg}
\affiliation{School of Physics and Astronomy, Raymond and Beverly
  Sackler Faculty of Exact Sciences, Tel Aviv University, Tel Aviv
  69978, Israel}
\author{M.~C.~Cross}
\affiliation{Department of Physics 114-36, California Institute of
  Technology, Pasadena, California 91125}
\author{Ron Lifshitz}
\email[Corresponding author:\ ]{ronlif@tau.ac.il}
\affiliation{School of Physics and Astronomy, Raymond and Beverly
  Sackler Faculty of Exact Sciences, Tel Aviv University, Tel Aviv
  69978, Israel}

\date{August 14, 2005}

\begin{abstract}
  We study the response of a large array of coupled nonlinear
  oscillators to parametric excitation, motivated by the growing
  interest in the nonlinear dynamics of microelectromechanical and
  nanoelectromechanical systems (MEMS and NEMS). Using a multiscale
  analysis, we derive an amplitude equation that captures the slow
  dynamics of the coupled oscillators just above the onset of
  parametric oscillations. The amplitude equation that we derive here
  from first principles exhibits a wavenumber dependent bifurcation
  similar in character to the behavior known to exist in fluids
  undergoing the Faraday wave instability. We confirm this behavior
  numerically and make suggestions for testing it experimentally with
  MEMS and NEMS resonators.
\end{abstract}

\maketitle

\section{Motivation}
In the last decade we have witnessed exciting technological advances
in the fabrication and control of microelectromechanical and
nanoelectromechanical systems (MEMS and NEMS).  Such systems are being
developed for a host of nanotechnological applications, as well as for
basic research in the mesoscopic physics of phonons, and the general
study of the behavior of mechanical degrees of freedom at the
interface between the quantum and the classical
worlds~\cite{R01,C03,blencowe}.  Surprisingly, NEMS have also opened
up a new experimental window into the study of the nonlinear dynamics
of discrete systems with many degrees of freedom. A combination of
three properties of NEMS resonators has led to this unique
experimental opportunity. First and most important is the experimental
observation that micro- and nanomechanical resonators tend to behave
nonlinearly at very modest amplitudes. This nonlinear behavior has not
only been observed
experimentally~\cite{turner98,C00,BR01,blick02,turner02,turner03,yu02,cleland04},
but has already been exploited to achieve mechanical signal
amplification and mechanical noise squeezing~\cite{rugar,carr} in
single resonators.  Second is the fact that at their dimensions, the
normal frequencies of nanomechanical resonators are extremely
high---recently exceeding the 1GHz
mark~\cite{HZMR03,cleland:070501}---facilitating the design of
ultra-fast mechanical devices, and making the waiting times for
unwanted transients bearable on experimental time scales.  Third is
the technological ability to fabricate large arrays of MEMS and NEMS
resonators whose collective response might be useful for signal
enhancement and noise reduction~\cite{sync}, as well as for
sophisticated mechanical signal processing applications. Such arrays
have already exhibited interesting nonlinear dynamics ranging from the
formation of extended patterns~\cite{BR02}---as one commonly observes
in analogous continuous systems such as Faraday waves---to that of
intrinsically localized modes~\cite{SHSICC03,SHSICC03b,SHSIC04}. Thus,
nanomechanical resonator arrays are perfect for testing dynamical
theories of discrete nonlinear systems with many degrees of freedom.
At the same time, the theoretical understanding of such systems may
prove useful for future nanotechnological applications.

%MCC This paragraph is new
Two of us (Lifshitz and Cross~\cite[henceforth LC]{LC03}) have
recently studied the response of coupled nonlinear oscillators to
parametric excitation. We used secular perturbation theory to
convert the equations of motion for the oscillators into a set of
coupled nonlinear \textit{algebraic\/} equations for the normal
mode amplitudes of the system, enabling us to obtain exact results
for small arrays but only a qualitative understanding of the
dynamics of large arrays. In order to obtain analytical results
for large arrays we study here the same system of equations,
approaching it from the continuous limit of infinitely-many
degrees of freedom. A novel feature of the paramerically driven
instability is that the bifurcation to standing waves switches
from supercritical (second order) to subcritical (first order) at
a wave number at or close to the critical one for which the
required driving force is minimum. This changes the form of the
amplitude equation that describes the onset of the paramerically
driven waves so that it no longer has the standard
\textquotedblleft Ginzburg-Landau\textquotedblright\ form. Our
central result is this new scaled amplitude
equation~(\ref{BampEq}), governed by a single control parameter,
that captures the slow dynamics of the coupled oscillators just
above the onset of parametric oscillations including this unusual
bifurcation behavior. We confirm the behavior numerically and make
suggestions for testing it experimentally. Although our focus is
on parametrically driven oscillators, the amplitude equation we
derive should also apply to other parametrically driven wave
systems with weak nonlinear damping.

\section{Equations of Motion}

%MCC This paragraph is changed
The equations of motion introduced in LC are
\begin{align}
\ddot{u}_{n}  & +u_{n}+u_{n}^{3}-\frac{1}{2}\Gamma(\dot{u}_{n+1}-2\dot{u}%
_{n}+\dot{u}_{n-1})\nonumber\label{eom}\\
& +\frac{1}{2}\Delta^{2}\bigl[1+H\cos(2\omega_{p}t)\bigr](u_{n+1}%
-2u_{n}+u_{n-1})\nonumber\\
& -\frac{1}{2}\eta\bigl[(u_{n+1}-u_{n})^{2}(\dot{u}_{n+1}-\dot{u}%
_{n})\nonumber\\
& -(u_{n}-u_{n-1})^{2}(\dot{u}_{n}-\dot{u}_{n-1})\bigr]=0,
\end{align}
with $n=1\ldots N$, and fixed boundary conditions
$u_{0}=u_{N+1}=0$. The equations of motion~(\ref{eom}) are modeled
after the experiment of Buks and Roukes \cite{BR02}, who succeeded
in fabricating, exciting, and measuring the response to parametric
excitation of an array of 67 micromechanical resonating gold
beams. Detailed arguments for the choice of terms introduced into
the equations of motion are given by LC. The guiding principle is
to introduce only those terms that are essential for capturing the
physical behavior observed in the experiment. These include a
cubic nonlinear elastic restoring force (whose coefficient is
scaled to 1), a dc electrostatic nearest-neighbor coupling term
with a small ac component responsible for the parametric
excitation (with coefficients $\Delta^{2}$ and $\Delta^{2}H$
respectively), and linear as well as cubic nonlinear dissipation
terms (with coefficients $\Gamma$ and $\eta$ respectively). The
nonlinear contribution to the dissipation, though expected to be
small, is essential for the saturation of the amplitude of the
parametrically driven motion~\cite{LC03} and is important in the
behavior of the bifircation as a function of wave number. Both
dissipation terms are taken to be of a nearest neighbor form,
motivated by the experimental indication that most of the
dissipation comes from the electrostatic interaction between
neighboring beams. It should be noted that the effect of
introducing gradient-dependent dissipation terms instead of local
dissipation terms is merely to renormalize the bare dissipation
coefficients $\Gamma$ and $\eta$ to wavenumber-dependent
coefficients of the form $\Gamma\sin^{2}\left(  q/2\right)  $ and
$\eta\sin^{2}\left(  q/2\right) $ respectively, as will become
apparent below.

The dissipation of the system is assumed to be weak, which makes it
possible to excite the beams with relatively small driving
amplitudes. In such case the response of the beams is moderate,
justifying the description of the system with nonlinearities up to
cubic terms only. The weak dissipation can be parameterized by
introducing a small expansion parameter $\epsilon\ll1$, physically
defined by the linear dissipation coefficient
$\Gamma\equiv\epsilon\gamma$, with $\gamma$ of order one. The
driving amplitude is then expressed by $\Delta^2 H=\epsilon h$, with
$h$ of order one. We assume the system is excited in its first
instability tongue, i.e. we take $\omega_p$ itself to lie within the
normal frequency band $\sqrt{1-2\Delta^2}<\omega_p<1$. The weakly
nonlinear regime is studied by expanding the displacements $u_n$ in
powers of $\epsilon$. Taking the leading term to be of the order of
$\epsilon^{1/2}$ ensures that all the corrections, to a simple set
of equations describing $N$ coupled harmonic oscillators, enter the
equations at the same order of $\epsilon^{3/2}$.

\section{Amplitude Equations for Counter Propagating Waves}

We introduce a continuous displacement field $u(x,t)$, keeping in
mind that only for integral values $x=n$ of the spatial coordinate
does it actually correspond to the displacements $u(n,t)=u_n(t)$ of
the discrete set of oscillators in the array. We introduce slow
spatial and temporal scales, $X=\epsilon x$ and $T=\epsilon t$, upon
which the dynamics of the envelope function occurs, and expand the
displacement field in terms of $\epsilon$,
\begin{eqnarray} \label{uAnsatz}\nonumber
u(x,t)&=& \epsilon^{1/2}
\bigl[\left(A_+(X,T)e^{-iq_px}+A_-^*(X,T)e^{iq_px}\right)e^{i\omega_p
t} \\
&+&  c.c.\bigr] + \epsilon^{3/2}u^{(1)}(x,t,X,T)+\ldots,
\end{eqnarray}
where the asterisk and $c.c.$ stand for the complex conjugate, and
$q_p$ and $\omega_p$ are related through the dispersion relation
$\omega_p^2=1-2\Delta^2\sin^2(q_p/2)$. The response to lowest order
in $\epsilon$ is expressed in terms of two counter-propagating waves
with complex amplitudes $A_+$ and $A_-$, a typical ansatz for
parametrically excited continuous systems~\cite{reviewcross}. We
substitute the ansatz~(\ref{uAnsatz}) into the equations of
motion~(\ref{eom}) term by term. Up to order $\epsilon^{3/2}$, we
have
\begin{widetext}
\begin{subequations}
\begin{align}
%%%%%%%%%%%%%%%%%%%%%%%%%%%%%%%%%%%%%%%%%%%%%%%%%%%%%%%%%
\ddot u_n&=-\epsilon^{1/2}\omega_p^2\left(A_+(X,T)e^{-iq_px}
+A_-^*(X,T)e^{iq_px}\right) e^{i\omega_pt}+\epsilon^{3/2}
\frac{\partial^2 u^{(1)}}{\partial
t^2}(x,t,X,T)\nonumber\\&+\epsilon^{3/2}2i\omega_p\left(\frac{\partial
A_+(X,T)}{\partial T}e^{-iq_px}+\frac{\partial A_-^*(X,T)}{\partial
T}e^{iq_px}\right)
e^{i\omega_pt}+ c.c.\,;\\
%%%%%%%%%%%%%%%%%%%%%%%%%%%%%%%%%%%%%%%%%%%%%%%%%%%%%%%%%%%%%%%%
u_{n\pm1}&=\epsilon^{1/2}
\left(A_+(X,T)e^{-iq_p(x\pm1)}+A_-^*(X,T)e^{iq_p(x\pm1)}\right)e^{i\omega_p
t}+\epsilon^{3/2}u^{(1)}(x\pm
1,t,X,T)\nonumber\\&\pm\epsilon^{3/2}\left(\frac{\partial
A_+}{\partial X}e^{-iq_px}e^{\mp iq_p}+\frac{\partial
A_-^*}{\partial X}e^{iq_px}e^{\pm
iq_p}\right)e^{i\omega_p t}+c.c.\,;\\
\displaybreak[0]
%%%%%%%%%%%%%%%%%%%%%%%%%%%%%%%%%%%%%%%%%%%%%%%%%%%%%%%%%%%%%%%%%%%%%%%%%%%
\frac12\Delta^2&\left(u_{n+1}-2u_n+u_{n-1}\right)=-\epsilon^{1/2}2\Delta^2
\sin^2(q_p/2)\left(A_+e^{-iq_px}+A_-^*e^{iq_px}\right)e^{i\omega_p
t}\nonumber\\&+\epsilon^{3/2}\left(u^{(1)}(x+1,t,X,T)-2u^{(1)}(x,t,X,T)+u^{(1)}(x-1,t,X,T)
\right)\nonumber\\&-\epsilon^{3/2}\Delta^2i\sin(q_p)\left(\frac{\partial
A_+}{\partial X}e^{-iq_px}-\frac{\partial A_-^*}{\partial
X}e^{iq_px}\right)e^{i\omega_p t}+c.c.\,;\\
\displaybreak[0]
%%%%%%%%%%%%%%%%%%%%%%%%%%%%%%%%%%%%%%%%%%%%%%%%%%%%%%%%%%%%%%%%%%%%%%%%%%%%%%%
\frac12\epsilon h{\rm c}&{\rm os}(2\omega_p
t)(u_{n+1}-2u_n+u_{n-1})=-\epsilon^{3/2}h\sin^2(q_p/2)\left(A_-e^{-iq_px}+A_+^*e^{iq_px}
\right)e^{i\omega_pt}+O(e^{i3\omega_pt})+c.c.\,;\\
%\nonumber\\
\displaybreak[0]
%%%%%%%%%%%%%%%%%%%%%%%%%%%%%%%%%%%%%%%%%%%%%%%%%%%%%%%%%%%%%%%%%%%%%%%%%%%%%%%
\frac12\epsilon\gamma(&\dot u_{n+1}-2\dot u_n+\dot
u_{n-1})=-\epsilon^{3/2}2i\gamma\omega_p\sin^2(q_p/2)\left(A_+e^{-iq_px}+A_-^*e^{iq_px}\right)e^{i\omega_p
t}+c.c.\,;
\\
%\nonumber\\
\displaybreak[0]
%%%%%%%%%%%%%%%%%%%%%%%%%%%%%%%%%%%%%%%%%%%%%%%%%%%%%%%%%%%%%%%%%%%%%%%%%%%%%%%
%%%%%%%%%%%%%%%%%%%%%%%%%%%%%%%%%%%%%%%%%%%%%%%%%%%%%%%%%%%%%%%%%%%%%%%%%%%%%%%
u_n^3=&\epsilon^{3/2}3\left[(|A_+|^2+2|A_-|^2)A_+e^{-iq_px}
+(2|A_+|^2+|A_-|^2)A_-^*e^{iq_px}\right]e^{i\omega_p
t}+O\left(e^{i3\omega_p t},e^{i3q_px}\right)+c.c.\,;\\
%\\ \nonumber
\displaybreak[0]
%%%%%%%%%%%%%%%%%%%%%%%%%%%%%%%%%%%%%%%%%%%%%%%%%%%%%%%%%%%%%%%%%%%%%%%%%%%%%%%%
\frac12\eta\left[\vphantom{\dot
u_n}\right.(&\left.u_{n+1}-u_n)^2(\dot u_{n+1}-\dot u_n)-
(u_n-u_{n-1})^2(\dot u_n-\dot u_{n-1})\right]=
-\epsilon^{3/2}i8\eta\omega_p\sin^4(q_p/2)\nonumber\\&\times\left[(|A_+|^2+2|A_-|^2)A_+
e^{-iq_px}+(2|A_+|^2+|A_-|^2)A_-^*
e^{iq_px}\right]e^{i\omega_pt}+O(e^{i3\omega_p
t},e^{i3q_px})+c.c.\,,
\end{align}
\end{subequations}
\end{widetext}
where $O(e^{i3\omega_p t},e^{i3q_px})$ are terms proportional to
$e^{i3\omega_p t}$ or $e^{i3q_px}$ which do not enter the dynamics at
the lowest order of the $\epsilon$ expansion. At the order of
$\epsilon^{1/2}$, the equations of motion~(\ref{eom}) are satisfied
trivially, yielding the dispersion relation mentioned earlier. At the
order of $\epsilon^{3/2}$ on the other hand we must apply a
\emph{solvability condition}~\cite{reviewcross}, which requires that
all terms proportional to $e^{(i\omega_pt\pm q_px)}$ must vanish. We
therefore obtain the two coupled amplitude equations,
\begin{eqnarray}\label{AmpEqs}\nonumber \frac{\partial
A_\pm}{\partial T} &\pm& v_g\frac{\partial A_{\pm}}{\partial X} = -
\gamma\sin^{2}\left(\frac{q_p}{2}\right)A_{\pm}
\mp i\frac{h}{2\omega_p}\sin^{2} \left(\frac{q_p}{2}\right)A_{\mp}\\
&-& \left(4\eta\sin^{4}\left(\frac{q_p}{2}\right) \mp
i\frac{3}{2\omega_p}\right)
\left(|A_{\pm}|^{2}+2|A_{\mp}|^{2}\right)A_{\pm},
\end{eqnarray}
where the upper signs (lower signs) give the equation for $A_+$
($A_-$) obtained from the restriction on the terms proportional to
$e^{(i\omega_pt-q_px)}$ ($e^{(i\omega_pt+q_px)}$), and
\begin{equation}
  \label{eq:vgroup}
  v_g =
\frac{\partial\omega}{\partial q} =
-\frac{\Delta^{2}\sin(q_p)}{2\omega_p}
\end{equation}
is the group velocity. A detailed derivation of the amplitude
equations~(\ref{AmpEqs}) can be found in Ref.~\onlinecite{yaron}.
Similar equations were previously derived for describing Faraday
waves~\cite{ezerskii,milner}.

\section{Reduction to a Single Amplitude Equation}
By linearizing the amplitude equations~(\ref{AmpEqs}) about the zero
solution ($A_+=A_-=0$) we find that the linear combination of the
two amplitudes that first becomes unstable at $h=2\gamma\omega_p$ is
$B\propto (A_+-iA_-)$---representing the emergence of a standing
wave with a temporal phase of $\pi/4$ relative to the drive---while
the orthogonal linear combination of the amplitudes decays
exponentially and does not participate in the dynamics at onset.
Thus, just above threshold we can reduce the description of the
dynamics to a single amplitude $B$, where at a finite amplitude
above threshold a band of unstable modes around $q_p$ can contribute
to the spatial form of $B$. This is similar to the procedure
introduced by Riecke~\cite{riecke} for describing the onset of
Faraday waves.

%MCC This paragraph is new
In the absence of nonlinear damping $\eta=0$ the coefficient of the
nonlinear term in Eq.~(\ref{AmpEqs}) is purely imaginary. This turns
out to yield the result that there is no nonlinear saturation term of
the usual form $\left\vert B\right\vert ^{2}B$ for the standing waves
at the resonant wave vector $q_{p}$. An analysis of the saturation of
standing waves of nearby wave numbers $q_{p}+k$ for this case shows
that the coefficient of $\left\vert B_{k}\right\vert ^{2}B_{k}$ is
positive for $k<0$, zero at $k=0$, and negative for $k>0.$This means
that for $\eta=0$ the nature of the bifurcation to standing waves
switches from supercritical to subcritical precisely at the critical
wave number $q=q_{p}$. Thus the amplitude equation describing the
onset of the standing waves will not be of the usual Ginzburg-Landau
form. It is the goal of the present section to derive the novel form
of the amplitude equation. We will do this for the case of $\eta$
small but non-zero, in which case the switch from subcritical to
supercritical occurs at a wave number close to the resonant wave
number $q_{p}$.

We define a reduced driving amplitude $g$ with respect to the
threshold $2\gamma\omega_p$ by letting
$(h-2\gamma\omega_p)/2\gamma\omega_p\equiv g\delta$, with $\delta\ll
1$.  In order to obtain an equation, describing the relevant slow
dynamics of the new amplitude $B$, we need to select the proper
scaling of the original amplitudes $A_\pm$, as well as their spatial
and temporal variables, with respect to the new small parameter
$\delta$. We assume that the coefficient of nonlinear dissipation
$\eta$ is small.
%MCC Changes
As we have seen, for some wave numbers near $q_{p}$ the bifurcation is then
subcritical, so that a quintic term must enter in order to saturate the growth
of the amplitude.
%It is thus apparent from the original amplitude
%equations~(\ref{AmpEqs}) that a quintic term must enter in order to
%saturate the growth of the amplitudes $A_\pm$.
This is similar to
the situation encountered by Deissler and Brand~\cite{deissler} who
studied localized modes near a subcritical bifurcation to traveling
waves. Here this can be achieved by defining the small parameter
$\delta$ with respect to the coefficient of nonlinear
dissipation---letting $\eta=\delta^{1/2}\eta_0$, with $\eta_0$ of
order one---and taking the amplitudes to be of order $\delta^{1/4}$.
Further noting that with a drive amplitude that scales as $\delta$
the growth rate scales like $\delta$ as well, and the bandwidth of
unstable wave numbers scales as $\delta^{1/2}$, we finally make the
ansatz that
\begin{eqnarray}\label{Bansatz}
\left(%
\begin{array}{c}
  A_+ \\
  A_- \\
\end{array}%
\right)&=&\delta^{1/4}
\left(%
\begin{array}{c}
  1 \\
  i \\
\end{array}%
\right) B(\hat\xi,\hat\tau)+
\delta^{3/4}\left(%
\begin{array}{c}
  w^{(1)}(X,T,\hat\xi,\hat\tau) \\
  v^{(1)}(X,T,\hat\xi,\hat\tau) \\
\end{array}%
\right)\nonumber\\&+&
\delta^{5/4}\left(%
\begin{array}{c}
  w^{(2)}(X,T,\hat\xi,\hat\tau) \\
  v^{(2)}(X,T,\hat\xi,\hat\tau) \\
\end{array}%
\right),
\end{eqnarray}
where $\hat\xi=\delta^{1/2} X$ and $\hat\tau=\delta T$ are the new
spatial and temporal scales respectively. The amplitude equation for
$B(\hat\xi,\hat\tau)$ is derived by once again using the multiple
scales method. We substitute the ansatz~(\ref{Bansatz}) into the
coupled amplitude equations~(\ref{AmpEqs}) and collect terms of
different orders of $\delta$.

%We express the terms of the coupled amplitude
%equations~(\ref{AmpEqs}) which with the ansatz~(\ref{Bansatz})
%become
%\begin{widetext}
%\begin{subequations}\label{deltacorrections}
%\begin{align}
%&|A_{+}|^{2}A_{+}=\delta^{3/4}|B|^{2}B+\delta^{5/4}(B^{2}{w^{(1)}}^*+2|B|^{2}w^{(1)})\,;\\&
%|A_{-}|^{2}A_{+}=\delta^{3/4}|B|^{2}B+\delta^{5/4}(iB^{2}{v^{(1)}}^*-i|B|^{2}v^{(1)}+|B|^{2}w^{(1)})\,;\\&
%|A_{+}|^{2}A_{-}=\delta^{3/4}|B|^{2}iB+\delta^{5/4}(iB^{2}{w^{(1)}}^*+i|B|^{2}w^{(1)}+|B|^{2}v^{(1)})\,;\\&
%|A_{-}|^{2}A_{-}=\delta^{3/4}|B|^{2}iB+\delta^{5/4}(-B^{2}{v^{(1)}}^*+2|B|^{2}v^{(1)})\,;\\&
%\frac{\partial A_{+}}{\partial T}=\delta^{3/4}\frac{\partial
%w^{(1)}}{\partial T}+\delta^{5/4}\frac{\partial w^{(2)}}{\partial
%T}+\delta^{5/4}\frac{\partial B}{\partial
%\hat\tau}\,;\\&\frac{\partial A_{-}}{\partial
%T}=\delta^{3/4}\frac{\partial v^{(1)}}{\partial
%T}+\delta^{5/4}\frac{\partial v^{(2)}}{\partial T}+\delta^{5/4}i
%\frac{\partial B}{\partial \hat\tau}\,;\\& \frac{\partial
%A_{+}}{\partial X}=\delta^{3/4}\frac{\partial w^{(1)}}{\partial
%X}+\delta^{3/4}\frac{\partial B}{\partial
%\hat\xi}+\delta^{5/4}\frac{\partial w^{(2)}}{\partial
%X}+\delta^{5/4}\frac{\partial
%w^{(1)}}{\partial\hat\xi}\,;\\&\frac{\partial A_{-}}{\partial
%X}=\delta^{3/4}\frac{\partial v^{(1)}}{\partial
%X}+\delta^{3/4}i\frac{\partial B}{\partial
%\hat\xi}+\delta^{5/4}\frac{\partial v^{(2)}}{\partial
%X}+\delta^{5/4}\frac{\partial v^{(1)}}{\partial\hat\xi}\,.
%\end{align}
%\end{subequations}
%\end{widetext}

Again, to the lowest order of expansion the equations are satisfied
trivially. Collecting all terms of order $\delta^{3/4}$ in
Eqs.~(\ref{AmpEqs}) yields
\begin{equation}\label{B3_4order}
\mathfrak{O}\left(%
\begin{array}{c}
  w^{(1)} \\
  v^{(1)} \\
\end{array}%
\right)=\left(-v_g\frac{\partial B}{\partial
\hat\xi}+i\frac9{2\omega_p}|B|^2B\right)
\left(%
\begin{array}{c}
  1 \\
  -i \\
\end{array}%
\right),
\end{equation}
where $\mathfrak{O}$ is a linear operator given by the matrix
\begin{equation} \label{B_Operator}
\left(%
\begin{array}{cc}
  \partial_T+v_g\partial_X+\gamma\sin^2(q_p/2) & i\gamma\sin^2(q_p/2) \\
  -i\gamma\sin^2(q_p/2) & \partial_T-v_g\partial_X+\gamma\sin^2(q_p/2) \\
\end{array}%
\right).
\end{equation}
There is no solvability condition to satisfy here because the right
hand side of Eq.~(\ref{B3_4order}) is automatically orthogonal to the
zero
eigenvalue of $\mathfrak{O}$, $\scriptsize\left(%
\begin{array}{c}
  1 \\
  i \\
\end{array}%
\right)$. The solution of Eq.~(\ref{B3_4order}) is given by
\begin{equation}\label{w1v1Sol}
\begin{split}
&\left(\begin{array}{c}
  w^{(1)} \\
  v^{(1)} \\
\end{array}%
\right)=\\&\frac1{2\gamma\sin^2(q_p/2)}\left(-v_g\frac{\partial
B}{\partial \hat\xi}+i\frac9{2\omega_p}|B|^2B\right) \left(
\begin{array}{c}
  1 \\
  -i \\
\end{array}%
\right).
\end{split}
\end{equation}
We plug Eq.~(\ref{w1v1Sol}) back into Eqs.~(\ref{AmpEqs}),
%using Eqs.~(\ref{deltacorrections})
collect all the terms of order $\delta^{5/4}$  and obtain

\begin{widetext}
\begin{multline}\label{B5_4order}
\mathfrak{O}\left(%
\begin{array}{c}
  w^{(2)} \\
  v^{(2)} \\
\end{array}%
\right)=\left[-\frac{\partial
B}{\partial\hat\tau}+\frac{v_g^2}{2\gamma\sin^2(q_p/2)}\frac{\partial^2
B}{\partial\hat\xi^2}+\gamma\sin^2(q_p/2)gB-12\eta_0\sin^4(q_p/2)|B|^2B\vphantom{1}\right.\\
+i\frac{3(-v_g)}{2\omega_p\gamma\sin^2(q_p/2)}\left(4|B|^2\frac{\partial
B}{\partial\hat\xi}+B^2\frac{\partial
B^*}{\partial\hat\xi}\right)-\frac{81}{8\omega_p^2\gamma\sin^2(q_p/2)}|B|^4B\left.\vphantom{\frac12}\right]
\left(%
\begin{array}{c}
  1 \\
  i \\
\end{array}%
\right).
\end{multline}
\end{widetext}
The solvability condition of Eq.~(\ref{B5_4order}), stating that the
zero
eigenvalue of the operator $\mathfrak{O}$, $\scriptsize\left(%
\begin{array}{c}
  1 \\
  i \\
\end{array}%
\right)$, must be orthogonal to the right hand side of the equation,
determines the amplitude equation for $B$~\cite{reviewcross}.
%% Therefore, $B$ must satisfy
%% \begin{multline}\label{BampEg}
%% \frac{\partial
%% B}{\partial\tau}=\gamma\sin^2(q_p/2)gB+\frac{v_g^2}{2\gamma\sin^2(q_p/2)}\frac{\partial^2
%% B}{\partial\xi^2}-12\eta_0\sin^4(q_p/2)|B|^2B\vphantom{1}\\
%% +i\frac{3|v_g|}{2\omega_p\gamma\sin^2(q_p/2)}\left(4|B|^2\frac{\partial
%% B}{\partial\xi}+B^2\frac{\partial B^*}{\partial\xi}\right)
%% -\frac{81}{8\omega_p^2\gamma\sin^2(q_p/2)}|B|^4B.
%% \end{multline}
Thus, the expression in square brackets in Eq.~(\ref{B5_4order}) must vanish.
After applying one last set of rescaling transformations,
\begin{eqnarray} \label{scaling}\nonumber
\hat\tau &=& \frac{9}{32} \frac{\tau}{\omega_p^2 \eta_0^2 \gamma
  \sin^{10}\left({q_p/2}\right)}, \quad
\hat\xi = \frac38 \frac{|v_g|\xi}{\omega_p \eta_0 \gamma
  \sin^6\left({q_p/2}\right)},\\
&&|B|^2 \rightarrow \frac{16}{27} \omega_p^2 \eta_0 \gamma
\sin^6\left(\frac{q_p}{2}\right) |B|^2,\nonumber\\&&g \rightarrow
\frac{32}{9} \omega_p^2 \eta_0^2
  \sin^8\left(\frac{q_p}{2}\right)g,
\end{eqnarray}
we end up with an amplitude equation governed by a single parameter,
\begin{eqnarray} \label{BampEq}\nonumber
\frac{\partial B}{\partial\tau} &=& gB +
\frac{\partial^{2}B}{\partial\xi^{2}} + i\frac{2}{3}
\left(4|B|^{2}\frac{\partial B}{\partial\xi} +
  B^{2}\frac{\partial B^{*}}{\partial\xi}\right)\\
&&-2|B|^{2}B - |B|^{4}B.
\end{eqnarray}

Note that Eq.~(\ref{BampEq}) is a scaled equation with all
coefficients of order unity. In an equation for the unscaled mode
amplitude $\tilde{B}=\delta ^{1/4}B$ (i.e.\ the combination of
$A_{+},A_{-}$ that goes unstable) and using the length and time
variables $X,T$ rather than $\xi,\tau$, the coefficient of the cubic
saturation term $|\tilde{B}|^{2}\tilde{B}$ would be small, scaling as
$\delta^{1/2}$, reflecting its origin in the nonlinear damping term,
the linear growth term proportional to $\tilde{B}$ would have a
coefficient scaling as $(h-h_{c})/h_{c}$, and all other coefficients
would be of order unity. All terms in the equation are then of
comparable size for $(h-h_{c})/h_{c}\sim\delta$, for oscillation
amplitudes $\tilde{B}\sim \delta^{1/4}$, and for length scales
$X\sim\delta^{-1/2}$ and time scales $T\sim\delta^{-1}$. The form of
Eq.~(\ref{BampEq}) is also applicable to the onset of parametrically
driven standing waves in continuum systems with weak nonlinear
damping, and combines in a single equation a number of effects studied
previously~\cite{ezerskii,milner,riecke,deissler,peilong00,peilong02}.
The numerical coefficients in the variable scalings and in the
equation will, of course, be different for different systems.

Once we have obtained this amplitude equation it can be used to study a
variety of dynamical solutions, ranging from simple single-mode to more
complicated nonlinear extended solutions, or possibly even localized solutions.

\begin{figure}
\includegraphics[width=0.9\columnwidth]{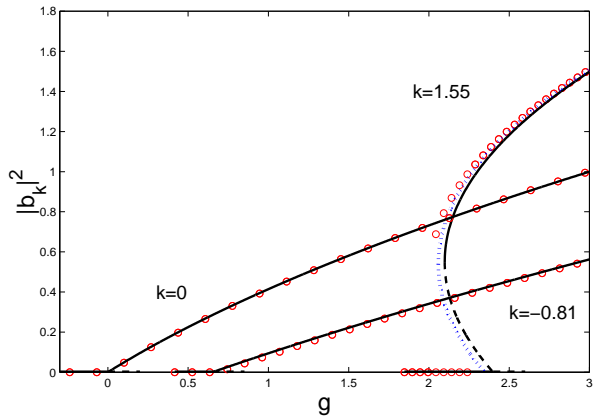}%
\caption{\label{NumericsFig} (Color online) Response of the oscillator
  array plotted as a function of reduced amplitude $g$ for three
  different scaled wave number shifts determined by fixing the number
  of oscillators $N$: $k=0$ (for $N=100$) and $k=-0.81$ (for $N=92$),
  which bifurcate supercritically, and $k=1.55$ (for $N=98$) which
  bifurcates subcritically showing clear hysteresis.  Solid and dashed
  lines are the positive and negative square root branches of the
  calculated response in~(\ref{Bsteady}), the latter clearly unstable.
  Open circles are numerical values obtained by integration of the
  equations of motion~(\ref{eom}) for each $N$, with $\Delta=0.5$,
  $\omega_p=0.767445$, $\epsilon\gamma=0.01$, and $\eta=0.1$. For
  these parameters a unity of the scaled squared amplitude $|b_k|^2$
  corresponds to $2\;10^{-4}$ for the unscaled squared amplitude, and
  a unity of the wave number shift $k$ corresponds to a shift of
  $9\;10^{-3}$ of the unscaled wavenumber. Dots show the
  subcritically-bifurcating single-mode solution of LC~\cite{LC03}.}
\end{figure}

\section{Single Mode Oscillations}

%MCC I've added a few words to provide better continuity
In this section we begin the investigation of the solutions of
Eq.~(\ref{BampEq}) by focusing on the regime of small reduced
drive amplitude $g$
%Here we focus on the regime of small reduced amplitude $g$
and look upon the saturation of single-mode solutions of the form
\begin{equation}\label{SingleMode}
B=b_ke^{-ik\xi},
\end{equation}
corresponding---via the scaling $\hat\xi=S_\xi \xi$, where the scale
factor $S_\xi$ is defined in~(\ref{scaling})---to a standing wave with
a shifted wave number $q=q_p + k \epsilon \sqrt{\delta}/S_\xi$.  From
the linear terms in the amplitude equation~(\ref{BampEq}) we find, as
expected, that for $g>k^2$ the zero-displacement solution is unstable
to small perturbations of the form of (\ref{SingleMode}), defining the
parabolic neutral stability curve, shown as a dashed line in
Fig.~\ref{StbBalloon}. The nonlinear gradients and the cubic term take
the simple form $2(k-1)|b_k|^2b_k$. For $k < 1$ these terms
immediately act to saturate the growth of the amplitude assisted by
the quintic term. Standing waves therefore bifurcate supercritically
from the zero-displacement state. For $k > 1$ the cubic terms act to
increase the growth of the amplitude, and saturation is achieved only
by the quintic term. Standing waves therefore bifurcate subcritically
from the zero-displacement state. Note that wave-number dependent
bifurcations similar in character were also predicted and observed
numerically in Faraday waves~\cite{milner,peilong00,peilong02}. The
saturated amplitude $|b_k|$, obtained by setting Eq.~(\ref{BampEq}) to
zero, is given by
\begin{equation}\label{Bsteady}
|b_k|^2 = (k-1) \pm \sqrt{(k-1)^2 + (g-k^2)}\geq 0.
\end{equation}
The original boundary conditions $u(0,t)=u(N+1,t)=0$ impose a phase
of $\pi/4$ on $b_k$ and require that the wave numbers be quantized
$q_m = m\pi/(N+1)$, $m=1\ldots  N$.

In Fig.~(\ref{NumericsFig}) we plot $|b_k|^2$ as a function of the
reduced driving amplitude $g$ for three different wave number shifts
$k$.  The solid (dashed) lines are the stable (unstable) solutions of
Eq.~(\ref{Bsteady}). The circles were obtained by numerical
integration of the equations of motion~(\ref{eom}). For each driving
amplitude, the Fourier components of the steady state solution were
computed to verify that only single modes are found, suggesting that
in this regime of parameters only these states are stable. By changing
the number $N$ of oscillators we could control the wave number shift
$k$ for a fixed value of $\omega_p$. In experiment it might be easier
to control $k$, for a fixed value of $\omega_p$, by changing the dc
component of the potential difference between the beams, thus changing
the dispersion relation and with it the value of $q_p$.

Lifshitz and Cross~\cite[Eq.~(33)]{LC03} obtained the exact form of
single mode solutions by substituting
$u_n=A_m\sin(q_mn)e^{i\omega_pt}+c.c$ directly into the equations of
motion~(\ref{eom}). In the limit of driving amplitudes just above
threshold and $\eta\ll1$ their solution corresponds to
Eq.~(\ref{Bsteady}), as shown by the dots in Fig.~(\ref{NumericsFig}).
In order to compare the two solutions one should note that in both
cases the system oscillates in one of its normal modes $q_m$ with the
driving frequency $\omega_p$. Here we use $k$ to denote the difference
between $q_p$ and the wavenumber of the oscillating pattern, whereas
Lifshitz and Cross use a frequency detuning $\omega_p-\omega_m$ to
denote the difference between the normal frequency and the actual
frequency of the oscillations. The frequency detuning
$\omega_p-\omega_m$ is proportional to $v_g k\epsilon \sqrt\delta$,
implying that for an infinitly extended system the standing waves will
always bifurcate supercritically with a wave number $q_p$ if the
driving amplitude is increased quasistatically. It is the discreteness
of the normal modes which provides the detuning essential for a
subcritical bifurcation if only quasistatic changes are performed.

\section{Secondary Instabilities}
\begin{figure}
\includegraphics[width=0.9\columnwidth]{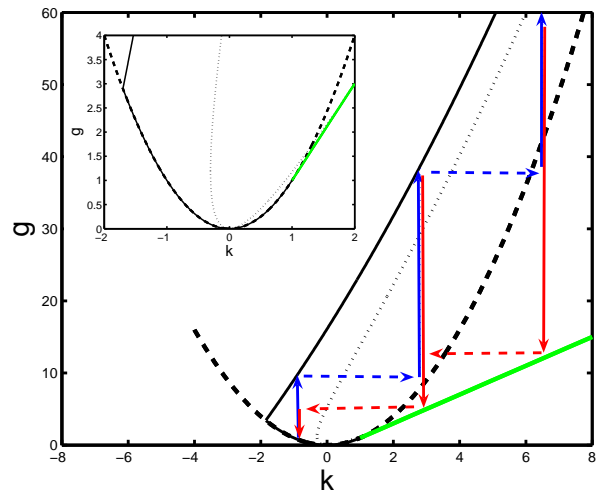}%
\caption{\label{StbBalloon}
  (Color online) Stability boundaries of the single-mode solution of
  Eq.~(\ref{BampEq}) in the $g$~{\it vs.}~$k$ plane.  Dashed line:
  neutral stability boundary. Dotted line: stability boundary of the
  single-mode solution~(\ref{SingleMode}) for a continuous spectrum.
  Solid line: stability boundary of the single-mode solution for
  $N=92$ and the parameters of Fig.~\ref{NumericsFig}.  For $k>1$ the
  bifurcation from zero displacement becomes subcritical and the lower
  stability boundary is the locus of saddle-node bifurcations (green
  line). Vertical and horizontal arrows mark the secondary instability
  transitions shown in Fig.~\ref{fig:SndIns}.}
\end{figure}

We study secondary instabilities of the single mode solutions by
performing linear stability analysis of~(\ref{SingleMode}). We
substitute
\begin{equation}\label{BStability}
B(\xi,\tau)=b_ke^{-ik\xi}+\left(\beta_+(\tau)
e^{-i(k+Q)\xi}+\beta_-^*(\tau)e^{-i(k-Q)\xi}\right),
\end{equation}
with $|\beta_\pm|\ll1$, into the amplitude equation~(\ref{BampEq})
and linearize in $\beta_\pm$. Since the amplitude
equation~(\ref{BampEq}) is invariant under phase transformations
$B\rightarrow Be^{-i\varphi}$, the stability of the single mode
solution cannot depend on the phase of $b_k$. We therefore assume
that $b_k$ is real, and linearize the following terms of
Eq.~(\ref{BampEq})
\begin{widetext}
\begin{eqnarray}
|B|^{2}B&\rightarrow&e^{-ik\xi}\left(b_{k}^{3}+b_{k}^{2}\left((2\beta_{+}+\beta_{-})e^{-iQ\xi}+(\beta_{+}^{*}+2\beta_{-}^{*})e^{iQ\xi}\right)\right)\nonumber\,;\\
|B|^{4}B&\rightarrow&e^{-ik\xi}\left(b_{k}^{5}+b_{k}^{4}\left((3\beta_{+}+2\beta_{-})e^{-iQ\xi}+(2\beta_{+}^{*}+3\beta_{-}^{*})e^{iQ\xi}\right)\right)\nonumber\,;\\
|B|^2\frac{\partial
B}{\partial\xi}&\rightarrow&e^{-ik\xi}\left(-ikb_{k}^{3}-ib_{k}^{2}\left(((2k+Q)\beta_{+}+k\beta_{-})e^{-iQ\xi}+
(k\beta_{+}^{*}+(2k-Q)\beta_{-}^{*})e^{iQ\xi}\right)\right)\nonumber\,;\\
B^{2}\frac{\partial
B^{*}}{\partial\xi}&\rightarrow&e^{-ik\xi}\left(ikb_{k}^{3}+ib_{k}^{2}\left((2k\beta_{+}+(k-Q)\beta_{-})e^{-iQ\xi}+
((k+Q)\beta_{+}^{*}+2k\beta_{-}^{*})e^{iQ\xi}\right)\right).\nonumber\\
\end{eqnarray}
The terms of order one of the equation obtained from the
linearization of Eq.~(\ref{BampEq}) recover the same
Eq.~(\ref{Bsteady}) for the single mode amplitude $b_k$. The terms
with spatial dependence of $e^{-iQ\xi}$ and $e^{iQ\xi}$ determine
the temporal development of the perturbations $\beta_+$ and
$\beta_-$ respectively, and are given by
\begin{eqnarray}
&&\frac\partial{\partial\tau}\left(
\begin{array}{c}
      \beta_+ \\
      \beta_- \\
\end{array}\right)=M\left(
\begin{array}{c}
      \beta_+ \\
      \beta_- \\
\end{array}
\right),\\&& M\equiv
\left(%
\begin{array}{cc}
 2b_k^2(k-1-b_k^2)-Q^2-2Q(k-4b_k^2/3) & 2b_k^2(k-1-b_k^2+Q/3) \\
 2b_k^2(k-1-b_k^2-Q/3) & 2b_k^2(k-1-b_k^2)-Q^2+2Q(k-4b_k^2/3) \\
\end{array}%
\right),\nonumber
\end{eqnarray}
\end{widetext}
where we have used Eq.~(\ref{Bsteady})
\begin{equation}
\begin{split}
g-k^2+4(k-1)b_k^2-3b_k^4=
%g-k^2+2(k-1)b_k^2-b_k^4+2b_k^2(k-1-b_k^2)=
2b_k^2(k-1-b_k^2).
\end{split}
\end{equation}
The single mode solution (\ref{SingleMode}) is a stable solution
only if all wave numbers $Q$ yield negative eigenvalues for $M$,
obtained for
\begin{eqnarray}\label{Mconditions}
trM&=&4b_k^2(k-1-b_k^2)-2Q^2<0\quad{\rm and}\\
|M|&=&\frac83Q^2\left(\frac38Q^2-b_k^4+\frac{5k+3}{2}b_k^2-\frac32k^2\right)>0\nonumber
\end{eqnarray}
for all wave numbers $Q$. The negative square root branch
in~(\ref{Bsteady}) (obtained for $b_k^2<k-1$) is confirmed to be
always unstable. The stability of the positive square root branch,
determined by setting the inequalities in Eqs.~(\ref{Mconditions})
to equalities, is bounded by the dotted curve in
Fig.~\ref{StbBalloon} describing the stability balloon of the single
mode state. Outside the stability balloon the standing wave
undergoes an Eckhaus instability with wave numbers $k\pm Q$, which
occurs first at $Q\rightarrow 0$. When taking into consideration the
discreteness of the system, where only wave numbers with $Q\geq
\Delta Q_N =  S_\xi \pi/\epsilon \sqrt{\delta} (N+1)$ can be taken
into account, the stability balloon is extended to the solid curve
in the Figure, plotted for the case of $N=92$.

\begin{figure}
\includegraphics[width=0.9\columnwidth]{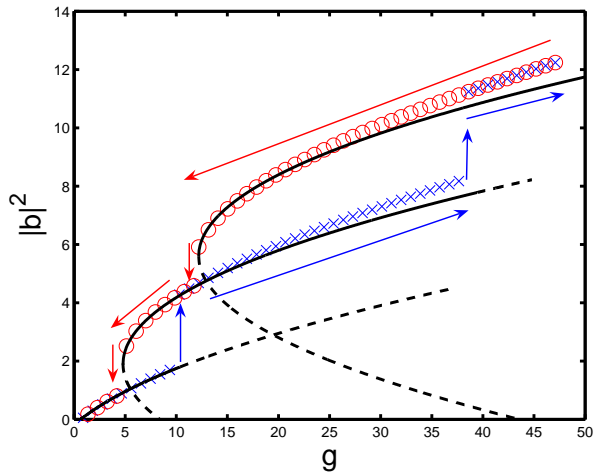}%
\caption{\label{fig:SndIns}
  (Color online) A sequence of secondary instabilities following the
  initial onset of single-mode oscillations in an array of $92$ beams,
  with the parameters of Fig.~\ref{NumericsFig}, plotted as a function
  of the reduced driving amplitude $g$. Solid (dashed) lines are
  stable (unstable) solutions defined by~(\ref{Bsteady}), for
  $k=-0.88$, $k'=2.81$, and $k''=6.51$, corresponding to the first
  wave number $k$ to emerge and two shifts to $k+\Delta Q_N$ and
  $k+2\Delta Q_N$ respectively.  Numerical integration of the
  equations of motion~(\ref{eom}) for an upward sweep of $g$ (blue
  $\times$'s), followed by a downward sweep (red $\circ$'s)
  exhibits clear hysteresis and confirms the theoretical predictions
  for the stability of single-mode oscillations as illustrated in
  Fig.~\ref{StbBalloon}.}
\end{figure}

A transition from one single-mode state to a new single-mode state
with a wave number shift of $n \Delta Q_N$, for some integer $n$,
occurs once the driving amplitude is increased and has crossed the
upper bound of the stability balloon. Since the upper bound
monotonically increases with $k$, the new wave number will always be
larger. A sequence of three transitions, obtained numerically, is
shown in Fig.~\ref{fig:SndIns}, superimposed with our theoretical
predictions.  The sequence of transitions is also sketched for
comparison within the stability balloon in Fig.~\ref{StbBalloon}.
This type of analysis yields predictions for hysteretic transitions on
slow sweeps, with results for when the transition occurs and for the
new state that develops, which for larger $g$ must be selected out of
a band of possible stable states.  It is clear that secondary
instabilities allow an additional scenario for observing subcritical
transitions, even in very large systems, where the wavenumber
separation is small. Once a secondary transition has occurred, the
system will return to oscillate in the original wavenumber only when
reducing the driving amplitude below the saddle node point of the
subcritical bifurcation.

\section{Conclusions}

%We have focused here on single-mode solutions of our newly-derived
%amplitude equation~(\ref{BampEq}) and their secondary instabilities,
%both of which have been numerically verified and should be
%experimentally tested on arrays of MEMS or NEMS resonators.  In the
%near future we intend to study the prediction of the amplitude
%equations for faster ramps of the driving amplitude, including the
%possibility of wavenumber jumps larger than $\Delta Q_N$ and
%multi-mode solutions arising from the nonlinear saturation of complex
%patterns growing from random initial displacements. In addition, we
%will investigate the behavior on slow and fast frequency sweeps.

We derived amplitude equations~(\ref{AmpEqs}) describing the
response of large arrays of nonlinear coupled oscillators to
parametric excitation, directly from the equations of motion
yielding exact expressions for all the coefficients. The dynamics at
the onset of oscillations was studied by reducing these two coupled
equations into a single scaled equation  governed by a single
control parameter~(\ref{BampEq}). Single mode standing waves were
found to be the initial states that develop just above threshold,
typical of parametric excitation. The single mode oscillations
bifurcate from the zero-displacement state either supercritically or
subcritically, depending on the wave number of the oscillations. The
wave number dependence originates from the nonlinear gradient terms of
the amplitude equation, which were usually disregarded in the past
by others in the analysis of parametric oscillations above
threshold. We also examined the stability of single mode
oscillations, predicting a transition of the initial standing wave
state to a new standing wave with a larger wave number as the
driving amplitude is increased.

In this work we showed that interesting response of coupled
nonlinear oscillators excited parametrically can also be obtained
for quasistatic driving amplitude sweeps, rather than the frequency
sweeps that are usually preformed in experiments.  We proposed, and
numerically demonstrated, two experimental schemes for observing our
predictions, hoping to draw more experimental attention  to the
dynamics produced by such sweeps of the driving amplitude.

The results obtained by the numerical integration of the equations of
motion agree with our analysis, supporting the validity of the
amplitude equation~(\ref{BampEq}). We therefore believe that the
amplitude equations we derived can serve as a good starting point for
studying other possible states of the system. One particular
interesting dynamical behavior that can be considered is that of
localized modes, often observed in arrays of coupled nonlinear
oscillators and in other nonlinear systems as well. The conditions for
obtaining such modes and their dynamical properties could be studied
by looking for localized states of the amplitude equations.  Another
interesting aspect that can be addressed using the amplitude equations
we derived is the response of the array to fast (rather than
quasistatic) sweeps of the driving amplitude, which should lead to
more complicated nonlinear response as observed by LC in their work.

\begin{acknowledgments}
  This research is supported by the U.S.-Israel Binational Science
  Foundation (BSF) under Grant No.~1999458, the National Science
  Foundation under Grant No.~DMR-0314069, and the PHYSBIO program with
  funds from the EU and NATO.
\end{acknowledgments}

\bibliography{bibpaper}

\end{document}